\documentclass[twocolumn]{aastex631}

\newcommand{\Msun}{\ensuremath{{\rm\;M_\odot}}}

\newcommand{\NEF}{NASA Einstein Fellow}

\begin{document}

\title{The First Photometric Evidence of a Transient/Variable Source at $z$\,$>$\,5 with \textit{JWST}}

\author[0000-0002-4781-9078]{Christa DeCoursey}
\affiliation{Steward Observatory, University of Arizona, 933 N. Cherry Ave, Tucson, AZ 85721, USA}

\author[0000-0003-1344-9475]{Eiichi Egami}
\affiliation{Steward Observatory, University of Arizona, 933 N. Cherry Ave, Tucson, AZ 85721, USA}

\author[0000-0002-4622-6617]{Fengwu Sun}
\affiliation{Center for Astrophysics $|$ Harvard \& Smithsonian, 60 Garden St., Cambridge MA 02138 USA}

\author[0009-0003-7532-3197]{Arshia Akhtarkavan}
\affiliation{Steward Observatory, University of Arizona, 933 N. Cherry Ave, Tucson, AZ 85721, USA}

\author[0000-0003-0883-2226]{Rachana Bhatawdekar}
\affiliation{European Space Agency (ESA), European Space Astronomy Centre (ESAC), Camino Bajo del Castillo s/n, 28692 Villanueva de la Cañada, Madrid, Spain}

\author[0000-0002-8651-9879] {Andrew J.\ Bunker}
\affiliation{Department of Physics, University of Oxford, Denys Wilkinson Building, Keble Road, Oxford OX1 3RH, UK}

\author[0000-0003-4263-2228]{David A. Coulter}
\affiliation{Space Telescope Science Institute, 3700 San Martin Drive, Baltimore, MD 21218, USA}

\author[0000-0003-0209-674X]{Michael Engesser}
\affiliation{Space Telescope Science Institute, 3700 San Martin Drive, Baltimore, MD 21218, USA}

\author[0000-0003-2238-1572]{Ori D. Fox}
\affiliation{Space Telescope Science Institute, 3700 San Martin Drive, Baltimore, MD 21218, USA}

\author[0000-0001-6395-6702]{Sebastian Gomez}
\affiliation{Department of Astronomy, The University of Texas at Austin, 2515 Speedway, Stop C1400, Austin, TX 78712, USA}

\author[0000-0001-9840-4959]{Kohei Inayoshi}
\affiliation{Kavli Institute for Astronomy and Astrophysics, Peking University, Beijing 100871, People's Republic of China}

\author[0000-0002-9280-7594]{Benjamin D.\ Johnson}
\affiliation{Center for Astrophysics $|$ Harvard \& Smithsonian, 60 Garden St., Cambridge MA 02138 USA}

\author[0000-0003-2495-8670]{Mitchell Karmen} 
\affiliation{Physics and Astronomy Department, Johns Hopkins University, Baltimore, MD 21218, USA}

\author[0000-0003-2037-4619]{Conor Larison}
\affiliation{Space Telescope Science Institute, 3700 San Martin Drive, Baltimore, MD 21218, USA}

\author[0000-0001-6052-4234]{Xiaojing Lin}
\affiliation{Department of Astronomy, Tsinghua University, Beijing 100084, China}
\affiliation{Steward Observatory, University of Arizona, 933 N. Cherry Ave, Tucson, AZ 85721, USA}

\author[0000-0002-6221-1829]{Jianwei Lyu}
\affiliation{Steward Observatory, University of Arizona, 933 N. Cherry Ave, Tucson, AZ 85721, USA}

\author[0000-0001-7497-2994] {Seppo Mattila}
\affiliation{Department of Physics and Astronomy, FI-20014 University of Turku, Finland}
\affiliation{School of Sciences, European University Cyprus, Diogenes Street, Engomi, 1516 Nicosia, Cyprus}

\author[0000-0003-1169-1954]{Takashi J. Moriya}
\affiliation{National Astronomical Observatory of Japan, National Institutes of Natural Sciences, 2-21-1 Osawa, Mitaka, Tokyo 181-8588, Japan}
\affiliation{Graduate Institute for Advanced Studies, SOKENDAI, 2-21-1 Osawa, Mitaka, Tokyo 181-8588, Japan}
\affiliation{School of Physics and Astronomy, Monash University, Clayton, Victoria 3800, Australia}

\author[0000-0002-2361-7201]{Justin D. R. Pierel} 
\altaffiliation{\NEF}
\affiliation{Space Telescope Science Institute, 3700 San Martin Drive, Baltimore, MD 21218, USA}

\author[0000-0001-8630-2031]{Dávid Puskás}
\affiliation{Kavli Institute for Cosmology, University of Cambridge, Madingley Road, Cambridge CB3 0HA, UK}
\affiliation{Cavendish Laboratory, University of Cambridge, 19 JJ Thomson Avenue, Cambridge CB3 0HE, UK}

\author[0000-0002-4410-5387]{Armin Rest}
\affiliation{Space Telescope Science Institute, 3700 San Martin Drive, Baltimore, MD 21218, USA}
\affiliation{Physics and Astronomy Department, Johns Hopkins University, Baltimore, MD 21218, USA}

\author[0000-0003-2303-6519]{George H. Rieke}
\affiliation{Steward Observatory, University of Arizona, 933 N. Cherry Ave, Tucson, AZ 85721, USA}

\author[0000-0002-4271-0364] {Brant Robertson}
\affiliation{Department of Astronomy and Astrophysics, University of California, Santa Cruz, 1156 High Street, Santa Cruz CA 96054, USA}

\author[0009-0000-0397-7894]{Sepehr Salamat}
\affiliation{Steward Observatory, University of Arizona, 933 N. Cherry Ave, Tucson, AZ 85721, USA}

\author[0000-0003-2238-1572]{Louis-Gregory Strolger} 
\affiliation{Space Telescope Science Institute, 3700 San Martin Drive, Baltimore, MD 21218, USA}

\author[0000-0002-8224-4505] {Sandro Tacchella}
\affiliation{Kavli Institute for Cosmology, University of Cambridge, Madingley Road, Cambridge CB3 0HA, UK}
\affiliation{Cavendish Laboratory, University of Cambridge, 19 JJ Thomson Avenue, Cambridge CB3 0HE, UK}

\author[0000-0001-5517-6335] {Christian Vassallo}
\affiliation{Department of Physics and Astronomy, FI-20014 University of Turku, Finland}

\author[0000-0003-2919-7495] {Christina C. Williams}
\affiliation{NSF's National Optical-Infrared Astronomy Research Laboratory, 950 North Cherry Ave, Tucson, AZ 85719, USA}

\author[0000-0002-0632-8897]{Yossef Zenati}
\affiliation{Physics and Astronomy Department, Johns Hopkins University, Baltimore, MD 21218, USA}
\affiliation{Space Telescope Science Institute, 3700 San Martin Drive, Baltimore, MD 21218, USA}
\affiliation{Astrophysics Research Center of the Open University (ARCO), The Open University of Israel, Ra'anana 4353701, Israel}

\author[0000-0002-1574-2045]{Junyu Zhang}
\affiliation{Steward Observatory, University of Arizona, 933 N. Cherry Ave, Tucson, AZ 85721, USA}

\correspondingauthor{Christa DeCoursey}
\email{cndecoursey@arizona.edu}

\begin{abstract}
The \textit{James Webb Space Telescope} (\textit{JWST}) discovered 79 transients out to $z$\,$\sim$\,4.8 through the JADES Transient Survey (JTS), but the JTS did not find any $z$\,$>$\,5 transients. We present the first photometric evidence of a $z$\,$>$\,5 transient/variable source with \textit{JWST}. The source, AT 2023adya, resides in a $z_{\mathrm{spec}}$\,$=$\,5.274 galaxy in GOODS-N, which dimmed from $m_{\rm F356W}$\,$=$\,26.05$\pm$0.02~mag to 26.24$\pm$0.02~mag in the rest-frame optical over approximately two rest-frame months, producing a clear residual signal in the difference image ($m_{\rm F356W}$\,$=$\,28.01$\pm$0.17~mag; SN$_\mathrm{var}$\,$=$\,6.09) at the galaxy center. Shorter-wavelength bands (F090W/F115W) show no rest-frame ultraviolet brightness change. Based on its rest-frame V-band absolute magnitude (M$_\mathrm{V}$\,$=$\,$-$18.48~mag), AT 2023adya could be any core-collapse supernova (SN) subtype or an SN Ia. However, due to low SN Ia rates at high redshift, the SN Ia scenario is unlikely. Alternatively, AT 2023adya may be a variable active galactic nucleus (AGN). The NIRCam/Grism spectrum shows no broad H$\alpha$ emission line (FWHM\,$=$\,130$\pm$26 km s$^{-1}$), but we cannot exclude the existence of a faint broad line and therefore we cannot exclude the AGN scenario. AT 2023adya is unlikely to be a tidal disruption event (TDE) because the TDE models matching the observed brightness changes have low event rates. Although it is not possible to determine  AT 2023adya's nature based on the two-epoch single-band photometry alone, this discovery pushes the transient/variable science frontier past $z$\,$=$\,5 and towards the epoch of reionization.
\end{abstract}  

\keywords{supernovae: general; galaxies: active}


\section{Introduction} 
\label{sec:intro}

With the launch of the \textit{James Webb Space Telescope} (\textit{JWST}), the redshift frontier of transient and variable science has extended far beyond the $z$\,$\sim$\,2 \textit{Hubble Space Telescope} (\textit{HST}) frontier. The high-redshift ($z$\,$\geq$\,2) transient and variable sources discovered with \textit{JWST} thus far include Type Ia supernovae (SNe Ia; the thermonuclear explosions of white dwarfs), core-collapse (CC) SNe (explosions of $\geq$8M$_\odot$ stars), and active galactic nuclei (AGN). 

Building a sample of high-redshift SNe Ia can provide useful constraints on SN Ia systematics \citep{Riess2006}, including whether they are truly similar to low-redshift SNe Ia and can act as standard candles \citep[]{Pierel2024a, Pierel2025}. This has important implications for cosmology constraints achieved at higher redshifts, such as the possibility of an evolving dark energy equation of state parameter,  w \citep{Adame2025}. CCSNe, on the other hand, closely trace the instantaneous star formation rate and thus provide an independent view of the cosmic star formation rate density. Observed CCSN rates can be compared to those expected from the cosmic star formation rate density to provide constraints on missing CCSN fractions, the progenitor mass range of CCSNe, and the potential scenario of an evolving initial mass function (IMF; \citealt{Dahlen2012, Strolger2015}). Additionally, there is evidence that CCSN explosion energies evolve with redshift \citep[]{Coulter2025, Moriya2025}. Building a larger sample of high-redshift CCSNe will generate better constraints on how high-redshift CCSNe differ from local CCSNe.

The \textit{JWST} Advanced Deep Extragalactic Survey (JADES) program obtained two epochs of deep ($\sim$30 mag) \textit{JWST} near-infrared camera (NIRCam) images of the Great Observatories Origins Deep Survey South (GOODS-S) field, separated by one observer-frame year \citep{Eisenstein2023}. The JADES Transient Survey (JTS) discovered 38 SNe at $z$\,$>$\,2, and the highest-redshift SN candidate in the JTS sample is associated with a $z_{\mathrm{phot}}$\,$=$\,4.82 host \citep{DeCoursey2025}. This sample includes an SN Ia at $z_\mathrm{spec}$\,$=$\,2.9 \citep{Pierel2024a}, a Type Ic broad line (Ic-BL) SN at $z_\mathrm{spec}$\,$=$\,2.83 \citep{Siebert2024}, and a likely Type IIP SN at $z_\mathrm{spec}$\,$=$\,3.61 \citep{Coulter2025}. \citet{Moriya2025} explores the properties of the robustly-classified Type II SNe in the JTS sample, three of which are at $z$\,$>$\,2. Additionally, within the COSMOS-Web SN survey, an SN Ia has been discovered at $z_\mathrm{spec}$\,$=$\,2.15 \citep{Pierel2025}. There have also been multiple serendipitous SN discoveries with \textit{JWST} \citep[e.g.][]{Yan2023}, including multiply-lensed SNe Ia at $z_\mathrm{spec}$\,$=$\,1.78 \citep[]{Frye2024, Pierel2024b, Pascale2025} and $z_\mathrm{spec}$\,$=$\,1.95 \citep{Pierel2024c}.

We conducted a systematic transient and variable search in the GOODS-North field with two sets of overlapping F090W, F115W, and F356W \textit{JWST}/NIRCam imaging, separated by one observer-frame year \citep{DeCoursey2024}. This search yielded a sample of 39 transient and variable sources, one of which is AT 2023adya. AT 2023adya is associated with a $z_{\mathrm{spec}}$\,$=$\,5.274 host galaxy \citep{SunF2024}, making it the highest-redshift transient/variable source showing a photometric brightness change discovered thus far with \textit{JWST}. We explore the types of transient or variable sources that may explain AT 2023adya, which include CCSNe, SNe Ia, variable AGN, and tidal disruption events (TDEs).


In Section \ref{sec:obs}, we describe the NIRCam observations covering AT 2023adya. Section \ref{sec:results} explains how we discovered AT 2023adya and presents the photometric measurements of AT 2023adya. We also describe AT 2023adya's host in Section \ref{sec:results}. In Section \ref{sec:discussion}, we step through the possible sources of AT 2023adya's transient or variable nature, and we summarize our findings in Section \ref{sec:conclusion}. Throughout this paper, we express magnitudes using the AB system \citep{OkeGunn1983} and adopt a flat $\Lambda$CDM cosmology with the following parameters: H$_0$\,$=$\,70 km s$^{-1}$ Mpc$^{-1}$, $\Omega_\Lambda$\,$=$\,0.7, and $\Omega_m$\,$=$\,0.3.


\section{Data} \label{sec:obs}

\subsection{NIRCam Imaging} \label{subsec:nircam_images}

The JADES program (program ID: 1181; PI: Eisenstein) obtained 9-band \textit{JWST}/NIRCam imaging of GOODS-N on UT 2023 February 3, using 4 short-wavelength (SW) wide-band filters (F090W, F115W, F150W, F200W), 3 long-wavelength (LW) wide-band filters (F277W, F356W, F444W), and 2 LW medium-band filters (F335M, F410M). The JADES observing strategy is described in detail in \citet{Eisenstein2023}, and the JADES images are publicly available on MAST/high-level science product (HLSP; \citealt{jades2024}), where AT 2023adya's host has ID 1026307.

Approximately one observer-frame year later on UT 2024 February 12--18, the Complete NIRCam Grism Redshift Survey (CONGRESS) program (program ID: 3577; PI: Egami) obtained 3-band \textit{JWST}/NIRCam imaging of GOODS-N, using two SW wide-band filters (F090W, F115W) and one LW wide-band filter (F356W) \citep{Egami2023}. The CONGRESS data is publicly available on MAST \citep{congress2025}. \citet{Rieke2023} provides a detailed description of the data processing procedure for JADES NIRCam observations of GOODS-S. Similar data processing procedures were applied to the JADES GOODS-N NIRCam observations as well as the CONGRESS NIRCam observations (F. Sun et al. in preparation).

We focus our analysis on the overlapping region in GOODS-N ($\sim$38 arcmin$^2$) between the F090W, F115W, and F356W JADES and CONGRESS images. Each image was mosaicked into the same pixel grid, with the F356W images resampled from their 0\farcs06/pixel native pixel scale to a 0\farcs03/pixel scale. The JADES images are hereafter referred to as Epoch1, and the CONGRESS images are hereafter referred to as Epoch2. 

\subsection{NIRCam/Grism Spectroscopy} \label{subsec:nircam_grism}

In addition to NIRCam imaging observations, CONGRESS simultaneously obtained $\sim$2-hr NIRCam/Grism Wide Field Slitless Spectroscopy (WFSS) observations of GOODS-N using the F356W filter. Previously, the First Reionization Epoch Spectroscopically Complete Observations (FRESCO) program (program ID: 1895; PI: Oesch) obtained $\sim$2-hr NIRCam/Grism WFSS observations of GOODS-N using the F444W filter on UT 2023 February 11--13 along with F444W, F182M, and F210M NIRCam imaging observations \citep{Oesch2023, fresco2023}. The spectral resolution of the NIRCam WFSS mode is R\,$\sim$\,1600 at 2.4--5.0 $\mu$m.

NIRCam grism data processing was detailed by \citet{SunF2024} following the public codes and calibrations\footnote{\dataset[doi:10.5281/zenodo.14052875]{https://zenodo.org/records/14052875} \citep{sun2024}} made available by \citet{SunF2023}, including latest updates on tracing and wavelength calibration based on \textit{JWST} Cycle-1 and 2 data.


\section{Results} \label{sec:results}

\subsection{AT 2023adya Discovery} \label{subsec:discovery}

\begin{figure*}
    \centering
 {\includegraphics[width=15cm]{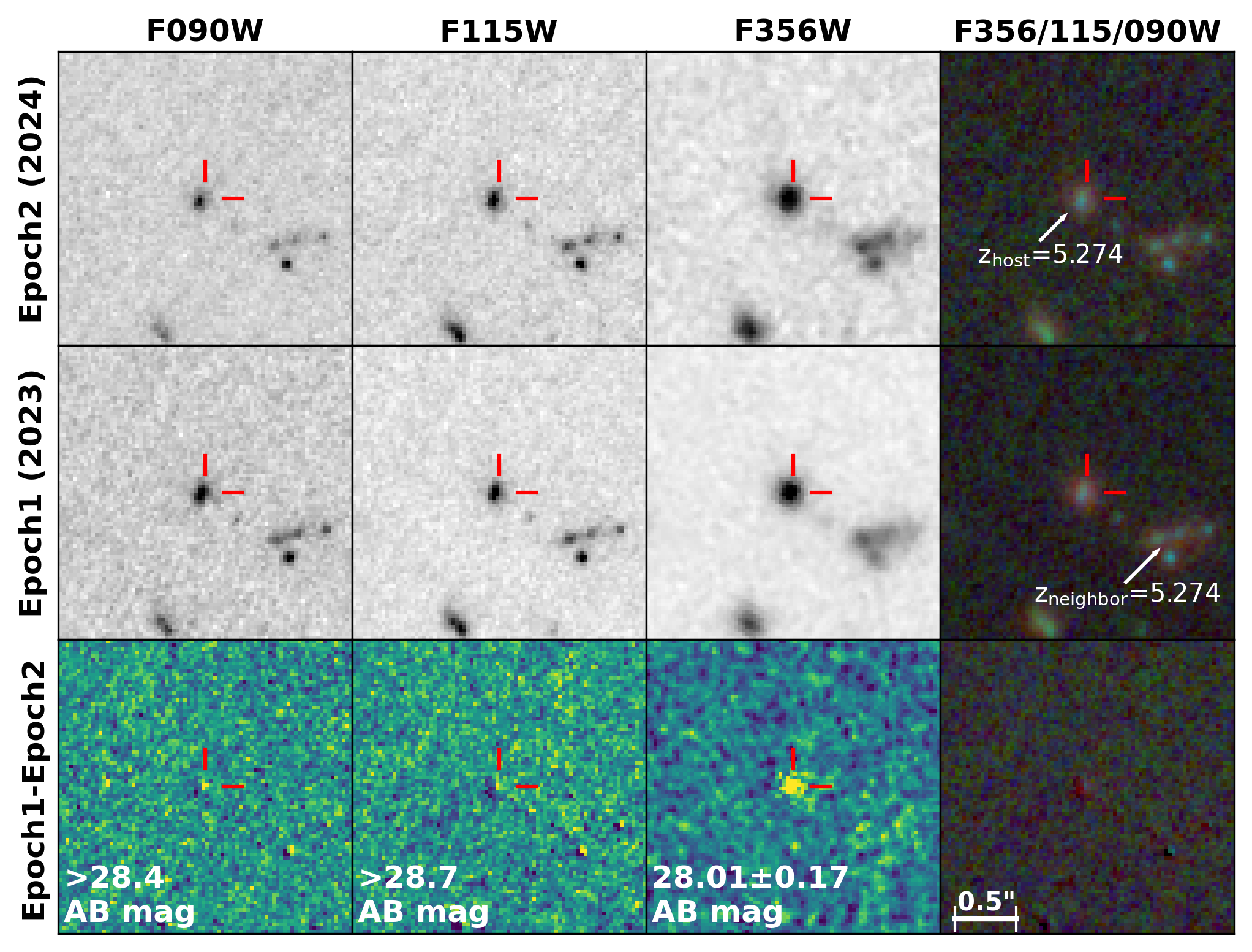}}
    \caption{NIRCam images of AT 2023adya, with red crosshairs indicating AT 2023adya's position. \textit{Top}: The Epoch2 (CONGRESS; 2024) F090W, F115W, and F356W NIRCam images. The right-most panel shows the F356W/F115W/F090W Epoch2 red-green-blue (RGB) image, with AT2023adya's host labeled. In these Epoch2 images, AT 2023adya has either faded or disappeared. \textit{Middle}: The Epoch1 (JADES; 2023) F090W, F115W, and F356W NIRCam images (and F356W/F115W/F090W RGB image) of AT 2023adya. AT 2023adya is present in the F356W image, although its emission is blended with its host's emission. The host's companion is labeled in the RGB image. \textit{Bottom}: The Epoch1-Epoch2 F090W, F115W, and F356W difference images (and F356W/F115W/F090W RGB difference image) showing AT 2023adya's change in brightness. There is clearly emission in the F356W difference image, indicating a change in brightness, but there is no emission in the F090W or F115W difference images.}
    \label{fig:stamp}
\end{figure*}

To search for transient and variable sources between Epoch1 and Epoch2, we created difference images for each filter by subtracting the Epoch2 data from the Epoch1 data. 
In addition to visually inspecting the difference images for point-like emission, we used \texttt{DAOSTarFinder} to search for evidence of transient/variable sources in each of the difference images \citep{Stetson1987}. We visually inspected the \texttt{DAOStarFinder} transient and variable candidates to remove contaminants such as noise spikes and diffraction spikes from the sample. This resulted in a sample of 39 transient and variable sources, which was reported to the Transient Name Server (TNS; \citealt{DeCoursey2024}). This sample includes AT 2023adya, positioned at (189\fdg12003, 62\fdg23867), which is exceptional because its host resides at $z_\mathrm{spec}$\,$=$\,5.274 \citep{SunF2024}. Figure \ref{fig:stamp} shows AT 2023adya in Epoch1, Epoch2, and the Epoch1-Epoch2 difference images in the F090W, F115W, and F356W filters.

\subsection{Photometry} \label{subsec:photometry_methods}

We performed aperture photometry using \texttt{astropy photutils} \citep{Bradley2024Astropy/photutils:1.12.0} on the difference image for each filter at AT 2023adya's position. We used an $r$\,$=$\,0\farcs2 circular aperture on the difference images with appropriate aperture corrections, and we used an $r$\,$=$\,0\farcs4--0\farcs6 annulus for sky subtraction. To measure flux uncertainty, we randomly placed $r$\,$=$\,0\farcs2 circular apertures in the background and calculated the standard deviation of the background flux. Refer to the third row of Table \ref{tab:phot} for the photometry measured from the difference images. We do not detect AT 2023adya in the F090W or F115W difference images, and hence $m_{\mathrm{F090W}}$\,$>$\,28.4~mag and $m_{\mathrm{F115W}}$\,$>$\,28.7~mag, which are 2$\sigma$ upper limits. However, the difference image photometry for F356W yields $m_{\mathrm{F356W}}$\,$=$\,28.01$\pm$0.17~mag. We list the rest-frame wavelengths at $z$\,$=$\,5.274 associated with each filter in the bottom row of Table \ref{tab:phot}.

\begin{deluxetable*}{cccc}[t]
\tablecaption{NIRCam Photometry \label{tab:phot}}   
\tablewidth{0pt}
\tablehead{
\colhead{} & \colhead{F090W} & \colhead{F115W} & \colhead{F356W}
}

\startdata
Epoch1                  & 26.54$\pm$0.05 & 26.51$\pm$0.04 & 26.05$\pm$0.02  \\
Epoch2                  & 26.48$\pm$0.06 & 26.45$\pm$0.05 & 26.24$\pm$0.02  \\
Epoch1$-$Epoch2           & $>$\,28.4 (2$\sigma$) & $>$\,28.7 (2$\sigma$) & 28.01$\pm$0.17  \\
SN$_{\mathrm{var}}$     & $<$0.80        & $<$0.92        & 6.09            \\
$\lambda_\mathrm{rest}$ & 144nm          & 183nm          & 567 nm          \\
\enddata
\tablecomments{The F356W difference image magnitude reported for AT 2023adya on TNS is slightly fainter ($m_{\mathrm{F356W}}$\,$=$\,28.27$\pm$0.09) due to the use of an 0\farcs1 aperture, which failed to capture all of AT 2023adya's emission.}
\end{deluxetable*}

To verify that the F356W difference image emission at AT 2023adya's position is real and not a subtraction artifact, we performed aperture photometry at AT 2023adya's position in the Epoch1 and Epoch2 science images in each filter and measured the brightness variation of the whole galaxy. The two epochs of science image photometry are listed in the top two rows of Table \ref{tab:phot}. The difference between the F356W Epoch1 and Epoch2 science image photometry yields nearly the exact same value as the photometry measured from the difference image ($m_{\mathrm{F356W}}$\,$=$\,28.04$\pm$0.14~mag), demonstrating that the F356W variability is real and not a subtraction artifact. AT 2023adya and its host were $\sim$19\% brighter in Epoch1 relative to Epoch2.

As an additional test to ensure the F356W brightness change is real, we randomly drew individual exposures covering AT 2023adya's position from the JADES and CONGRESS datasets to produce bootstrapped Epoch1 and Epoch2 mosaics. There were 8 (7) total exposures in the Epoch1 JADES (Epoch2 CONGRESS) data covering AT 2023adya. So, to create the bootstrapped mosaics, we randomly chose 7 (6) exposures and removed duplicates, resulting in 5--7 (4--6) exposures used to generate the bootstrapped Epoch1 JADES (Epoch2 CONGRESS) mosaics. We then performed aperture photometry on each bootstrapped mosaic and compared the results to the galaxy fluxes measured from the full JADES and CONGRESS mosaics (rows 1--2 of Table \ref{tab:phot}). We performed this exercise six times and found that the measured fluxes from the bootstrapped Epoch1 and Epoch2 mosaics were always consistent with the JADES and CONGRESS galaxy fluxes within 1$\sigma$, respectively. In other words, our bootstrapping test validates that the F356W brightness change is not an artifact due to certain bad exposures.

Additionally, we computed the signal-to-noise ratio of variability (SN$_{\mathrm{var}}$) of the science image photometry using the following equation: 

\begin{equation}
    \mathrm{SN_{var}} = \frac{|\mathrm{flux_{Epoch1}} - \mathrm{flux_{Epoch2}}|}{\sqrt{\sigma^2_{\mathrm{flux,Epoch1}} + \sigma^2_{\mathrm{flux,Epoch2}}}},
\end{equation}
where $\sigma_{\mathrm{flux}}$ is the flux uncertainty. The SN$_{\mathrm{var}}$ for F090W, F115W and F356W are $<$0.80, $<$0.92, and 6.09, respectively, further indicating that AT 2023adya exhibits a change in F356W brightness but does not exhibit a change in F090W or F115W brightness.

\subsection{AT 2023adya's Position and Host} \label{subsec:position}

AT 2023adya was discovered at (189\fdg12003, 62\fdg23867) in GOODS-N and belongs to host galaxy JADES--GN+189.12004+62.23867. The host resides at $z_{\mathrm{spec}}$\,$=$\,5.274 \citep[][the redshift is updated based on improved calibrations]{SunF2024}, measured unambiguously from an H$\alpha$ emission line in the FRESCO F444W NIRCam/Grism spectrum and an [\ion{O}{3}] $\lambda$5008 emission line in the CONGRESS F356W NIRCam/Grism spectrum. \citet{SunF2024} measured AT 2024adya's host to be centered at (189\fdg12004, +62\fdg23867), which is only $\sim$0\farcs02 ($\sim$0.10 kpc) away from AT 2023adya's position. We re-measured the host centroid using the Epoch2 F356W image since \citet{SunF2024} measured the centroid using the AT 2023adya-contaminated images (i.e., Epoch1 images), but we find a negligible difference in the Epoch2 host centroid. Hence, we use the measurement from \citet{SunF2024} as the host centroid. The angular separation between AT 2023adya and its host's center is $\sim$1/3 of a native LW pixel, and thus AT 2023adya can be considered coincident with the host's center within the uncertainty, as shown in Figure \ref{fig:stamp}. 

\begin{figure*}
    \centering
 {\includegraphics[width=15cm]{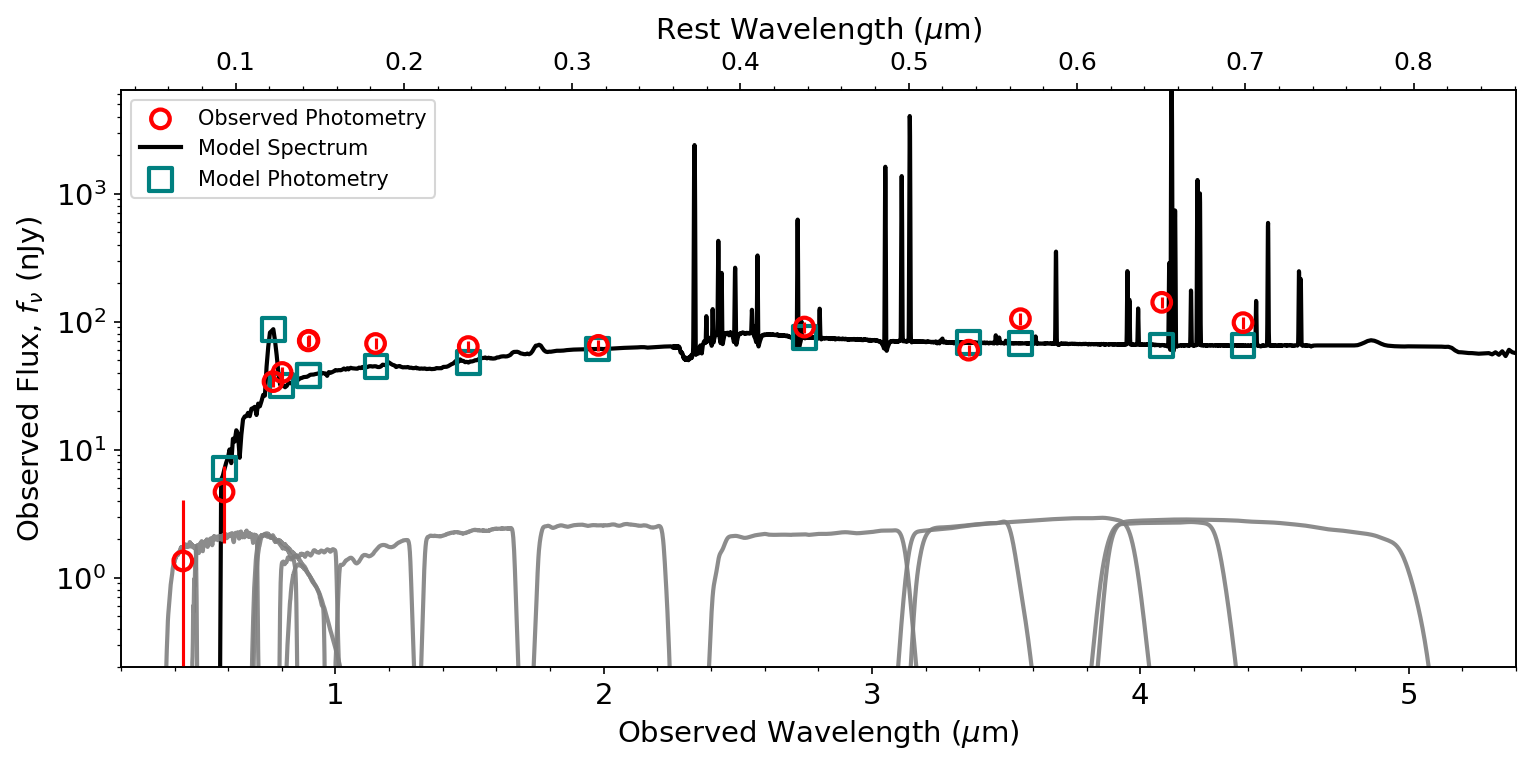}}
    \caption{AT 2023adya host's SED measured from the Epoch1 \textit{JWST}/NIRCam JADES photometry and \textit{HST}/ACS HLF photometry. The red circles show the observed photometry with uncertainties, the teal squares show the model photometry, the black curve shows the best-fit model spectrum, and the gray curves show the filter transmission curves. The model spectrum arises solely from stellar contributions, showing no evidence of an AGN presence.}
    \label{fig:host_sed}
\end{figure*}

Although the 9-band JADES \textit{JWST}/NIRCam imaging from Epoch1 is contaminated with AT 2023adya, we performed host galaxy spectral energy distrubtion (SED) fitting with a modified version of \texttt{Prospector} \citep{johnson2021, Lyu2022} to estimate the host properties and attempt to identify whether or not an AGN is present (Figure \ref{fig:host_sed}). The contamination from AT 2023adya is unlikely to significantly skew the fit because it represents $<$20\% of the F356W flux and does not affect the F090W or F115W flux. The contamination from AT 2023adya presumably affects the other LW filters to a similar extent as the $<$20\% contribution to the F356W flux, though we cannot confirm this. We used the cataloged JADES photometry \citep{jades2024} and \textit{HST} Advanced Camera for Surveys (ACS) photometry measured from the \textit{Hubble} Legacy Fields (HLF) data release v2.5 \citep{hlf2015} to perform the fit.

Although our photometric coverage is limited in wavelength, the best-fit SED does not rise toward longer wavelength, showing no obvious sign of AGN contribution to the total emission; the best-fit model spectrum arises solely from stellar contributions. Based on the best-fit model, the host has a mass of log(M/M$_\odot$)\,$=$\,9.13$^{+0.07}_{-0.11}$, a metallicity of log(Z/Z$_\odot$)\,$=$\,-1.82$^{+0.26}_{-0.13}$, and a V-band stellar dust attenuation of A$_\mathrm{V}$\,$\sim$\,0.2. The computed star formation rate is 7.8 M$_\odot$ year$^{-1}$ assuming a delayed tau star formation history with a stellar age at 0.27 Gyr and a characteristic timescale of $\tau$\,$=$\,0.10 Gyr.

Interestingly, AT 2023adya's host is a member of a high-confidence merger pair ($\sim$80\% close pair probability; \citealt{Puskas2025}), residing only $\sim$5.2 kpc from JADES-GN-189.11960+62.23855 ($z_{\mathrm{spec}}$\,$=$\,5.274; \citealt{SunF2024}), and is also part of overdensity JADES--GN--OD--5.269 \citep{Helton2024}.


\section{Discussion} \label{sec:discussion}

Because we only detect a source brightness change in one filter between two imaging epochs, we cannot conclusively determine AT 2023adya's origin. However, we examine various possibilities, including an SN explosion, merger-induced AGN variability, and a TDE.

\subsection{Supernova} \label{subsec:sn}


We explore a variety of SN subtypes at $z$\,$=$\,5.274 to determine which SN subtypes can mimic AT 2023adya's F356W magnitude. At $z$\,$=$\,5.274, $m_\mathrm{F356W}$\,$=$\,28.01~mag translates to a rest-frame 567 nm (approximately V-band) absolute magnitude of M$_\mathrm{V}$\,$=$\,$-$18.48~mag. This absolute magnitude provides constraints on the types of SNe that may explain AT 2023adya. 

\begin{figure}
    \centering
 {\includegraphics[width=8.5cm]{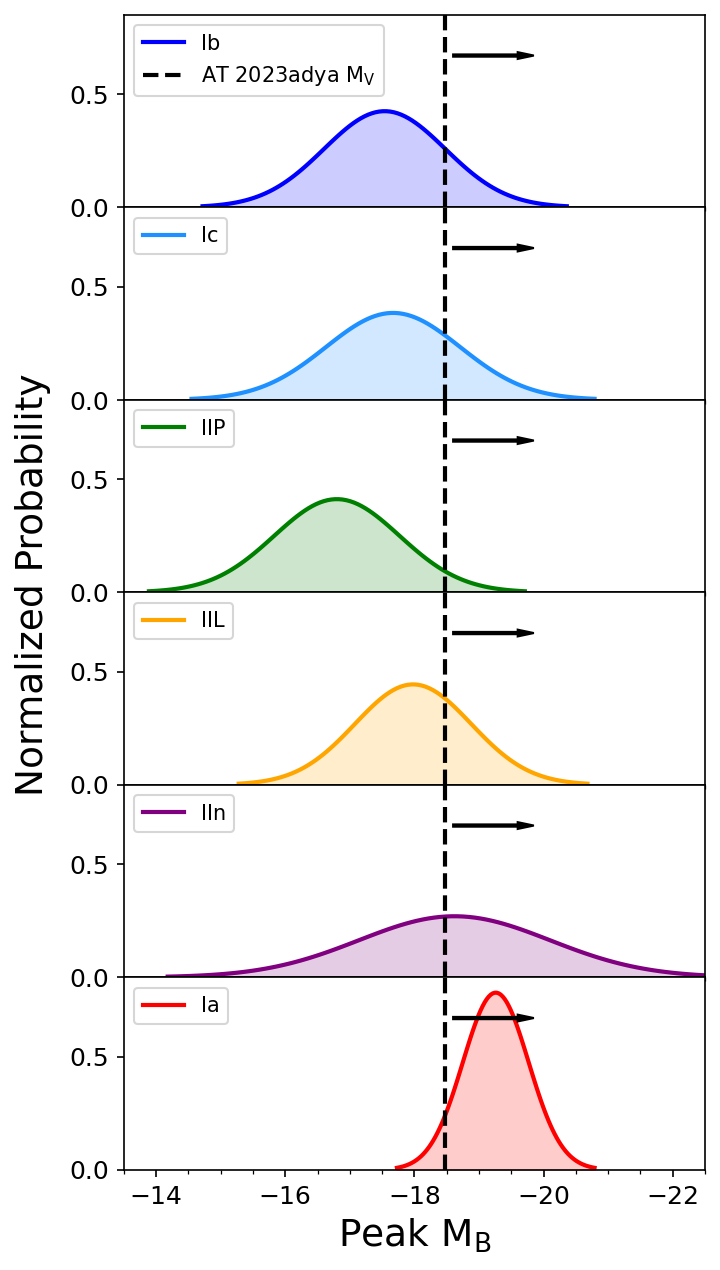}}
    \caption{The Gaussian distributions of peak absolute B-band magnitudes from \citet{Richardson2014} for SNe Ib (dark blue), Ic (light blue), IIP (green), IIL (orange), IIn (purple), and Ia (red). The vertical black dashed line shows AT 2023adya's rest-frame 567 nm (approximately V-band) absolute magnitude. It is unlikely that AT 2023adya was exactly at its V-band peak in Epoch1 and AT 2023adya may still be present in Epoch2, so AT 2023adya's V-band absolute magnitude can be treated as a lower limit on brightness relative to the peak absolute B-band magnitude distributions (hence the black arrows).}
    \label{fig:sne}
\end{figure}

We plot the volume-limited Gaussian distribution of peak absolute B-band magnitude for SNe Ib, Ic, IIP, IIL, IIn, and Ia in Figure \ref{fig:sne} \citep{Richardson2014}. The \citet{Richardson2014} absolute B-band magnitudes primarily come from the low-redshift Asiago Supernova Catalog sample. We will treat AT 2023adya's M$_\mathrm{V}$\,$=$\,$-$18.48~mag as a lower limit on brightness compared to the peak B-band distributions because (1) it is unlikely that we observed AT 2023adya at its peak V-band absolute magnitude, and (2) it is possible that AT 2023adya is still present in Epoch2, which would cause the difference image photometry to underestimate AT 2023adya's absolute brightness. 

This interpretation implies that only SNe which have absolute B-band magnitudes equal to or brighter than M$_\mathrm{V}$\,$=$\,$-$18.48~mag can explain AT 2023adya. However, it should be noted that SNe Ib/c peak brighter in the rest-frame V-band than B-band \citep[]{Woosley2021, Harim2023}, and SNe II and SNe Ia tend to peak at similar rest-frame V-band and B-band absolute magnitudes or slightly brighter in the rest-frame V-band than B-band \citep[]{Ashall2016, deJaeger2019, Harim2023, Anderson2024}. Therefore, treating AT 2023adya's absolute V-band magnitude as a lower limit on brightness relative to the peak B-band absolute magnitude distributions will result in conservative conclusions on the SN subtypes that are compatible with AT 2023adya, since the V-band peak absolute magnitude distributions are similar or brighter than those of the B-band. 

As shown in Figure \ref{fig:sne}, SNe Ib, Ic, and IIL with typical peak B-band absolute magnitudes are too faint to explain AT 2023adya, but the bright tails of these CCSN subtypes are sufficiently luminous. Both typical and bright SNe IIn and Ia are bright enough to explain AT 2023adya. 

\subsubsection{SN IIP}
At $z$\,$=$\,5.274, F115W approximately corresponds to the rest-frame near-UV (NUV; $\sim$183nm) and F356W approximately corresponds to the rest-frame V-band. As seen in Figure \ref{fig:sne}, only the brightest SNe IIP are sufficiently luminous to explain AT 2023adya. No typical low-redshift SN IIP is comparable to AT 2023adya. 

However, we can compare AT 2023adya to SN 2009kf, a $z$\,$=$\,0.182 SN IIP that was bright not only in V-band (M$_\mathrm{V, peak}$\,$\approx$\,$-$19.45~mag) but also incredibly bright in the NUV (M$_\mathrm{NUV,peak}$\,$\approx$\,$-$21.5$\pm$0.5~mag). Dense circumstellar medium interaction caused the unusual NUV brightness \citep{Botticella2010, Moriya2011}. 

At SN 2009kf's redshift of $z$\,$=$\,0.182, the effective \textit{Galaxy Evolution Explorer} (\textit{GALEX}) NUV wavelength corresponds to rest-frame $\sim$195nm and thus probes a similar rest-frame wavelength as AT 2023adya's F115W photometric non-detection. F115W's limiting magnitude of $m_\mathrm{F115W}$\,$\leq$\,28.7~mag corresonds to M$_\mathrm{NUV}$\,$\lesssim$\,$-$17.8~mag at $z$\,$=$\,5.274, and if AT 2023adya is similar to SN 2009kf, it would have peaked around M$_\mathrm{NUV}$\,$=$\,$-$21.5~mag, well above the detection limit. In other words, if AT 2023adya is SN 2009kf-like, it probably would have been detected in F115W. However, we cannot rule out the possibility of dust-extinction causing the NUV non-detection.

We are thus unable to rule out the possibility that AT 2023adya is a luminous, potentially dust-extincted SN IIP. The high-redshift SNe II from the JTS have tentatively shown evolution toward higher luminosities and hence toward higher explosion energies \citep[]{Coulter2025, Moriya2025}, highlighting the possibility that AT 2023adya is a luminous SN IIP.

\subsubsection{SN Ia}
Additionally, most SNe Ia, including underluminous 1991bg-like SNe Ia \citep[]{Filippenko1992, Graur2024}, are sufficiently bright to explain AT 2023adya. Only the least luminous SNe Ia are incompatible with AT 2023adya's F356W magnitude, as shown in the bottom panel of Figure \ref{fig:sne}. 

The SN Ia delay-time distribution (DTD) is the distribution of elapsed time between a hypothetical instantaneous burst of star formation and an SN Ia explosion. The observed DTD ranges from $\sim$10 Myr to $\sim$10 Gyr \citep[]{Maoz2012,GraurMaoz2013,MaozGraur2017,Liu2023}. Assuming that AT 2023adya is an SN Ia and the progenitor formed at $z$\,$\sim$\,10, the delay-time would be $\sim$0.6 Gyr, which is within the expected range of DTDs. It is therefore physically possible that AT 2023adya is an SN Ia. 

However, due to the decline of SN Ia rates at high redshift, it is unlikely that AT 2023adya is an SN Ia. \citet{Rodney2014} computed volumetric SN Ia rates out to $z$\,$\sim$\,2.5 based on \textit{HST} observations. Their Figure 9 hints at a decrease in the SN Ia rate beyond $z$\,$\sim$\,2, though the error bars are large. \citet{Palicio2024} computed theoretical SN Ia rates by convolving various observed cosmic star formation rates with various DTDs. The theoretical SN Ia rates shown in their Figure 3 show a strong decline beyond $z$\,$\sim$\,2 where the longer delay times become unphysical. Therefore, although the photometry we measure matches what we would expect for an SN Ia, the sheer number of CCSNe we would expect at this redshift compared to SNe Ia leads us to believe AT 2023adya is more likely of the former class.

\subsubsection{Exotic SNe}
To explore the superluminous SN (SLSN) possibility, we take the SLSN-I light curve model from \citet{Moriya2022} and redshift it to $z$\,$=$\,5.274. We find that this SLSN model is compatible with AT 2023adya's observational signatures if it is post-peak, as the F115W light curve will be below the detection threshold while the F356W light curve will be well above the detection threshold. However, given the rarity of SLSNe and that AT 2023adya is compatible with normal SNe, it is unlikely that AT 2023adya is an SLSN.

We also explore a variety of pair-instability SN (PISN) model light curves from \citet{Kasen2011} and found that the bare helium core and hydrogenic red supergiant models at a variety of masses can reproduce the F356W detection and lack of F090W and F115W detections at $z$\,$=$\,5.274. However, again given the rarity of PISNe and that AT 2023adya is compatible with many normal SNe, there is no compelling evidence that AT 2023adya is a PISN.

We cannot fully explore the fast blue optical transient (FBOT) possibility because we do not have rest-frame \textit{r}-band or \textit{g}-band measurements (selection generally requires \textit{g}--\textit{r}\,$\lesssim$\,-0.2; \citealt{Drout2014}), and we do not have the proper cadence to determine the rise and fall time (selection generally requires that the duration above the light curve half-maximum to be 1 day\,$<$t$_{1/2}$\,$<$\,12 days; \citealt{Ho2023}). However, \citet{Drout2014} estimates a volumetric FBOT rate of 4--7\% of the CCSN rate at $z$\,$\sim$\,0.2 \citep{Botticella2008}, and \citet{Ho2023} estimates that the luminous FBOT (LFBOT) rate is $\lesssim$\,0.1\% of the local CCSN rate, making it unlikely that AT 2023adya is an FBOT or LFBOT.


\subsection{Active Galactic Nucleus} \label{subsec:agn}

\subsubsection{Type 1 AGN} \label{subsubsec:type1_agn}

Optical variability is a common selection technique to identify AGN \citep[]{Klesman2007, Trevese2008, Villforth2010, DeCicco2015, Garcia-Gonzalez2015, Yuk2022}, as many Type 1 AGN exhibit variability on timescales ranging from hours to years \citep[]{Ulrich1997, VandenBerk2004, Sesar2007, Hickox2014}. Type 1 AGN exhibit a broad H$\alpha$ emission line ($\geq$\,1000 km/s) that cannot be attributed to normal star-forming galaxy activity. Under the widely-accepted assumption that this broad H$\alpha$ emission is connected to AGN activity, the AGN broad line region (BLR) generates the broad H$\alpha$ emission. 

There have been numerous searches for AGN variability with \textit{JWST}, especially with regards to Type 1 AGN and ``Little Red Dots" (LRDs: red, compact, high-redshift objects which are thought to be dusty AGN; \citealt[]{Kokubo2024, Zhang2024}).  Photometric variability has generally not yet been observed for Type 1 AGN and LRDs with the limited amount of \textit{JWST} data available so far. However, there is evidence of spectral variability in multiply-imaged LRD A2744-QSO1 at $z$\,$=$\,7.04 \citep[]{Ji2025, Furtak2025}. The JTS reports seven variable AGN candidates, none of which are LRDs or have been identified as Type 1 AGN \citep{DeCoursey2025}. 

\begin{figure*}
\centering
\includegraphics[width=0.49\linewidth]{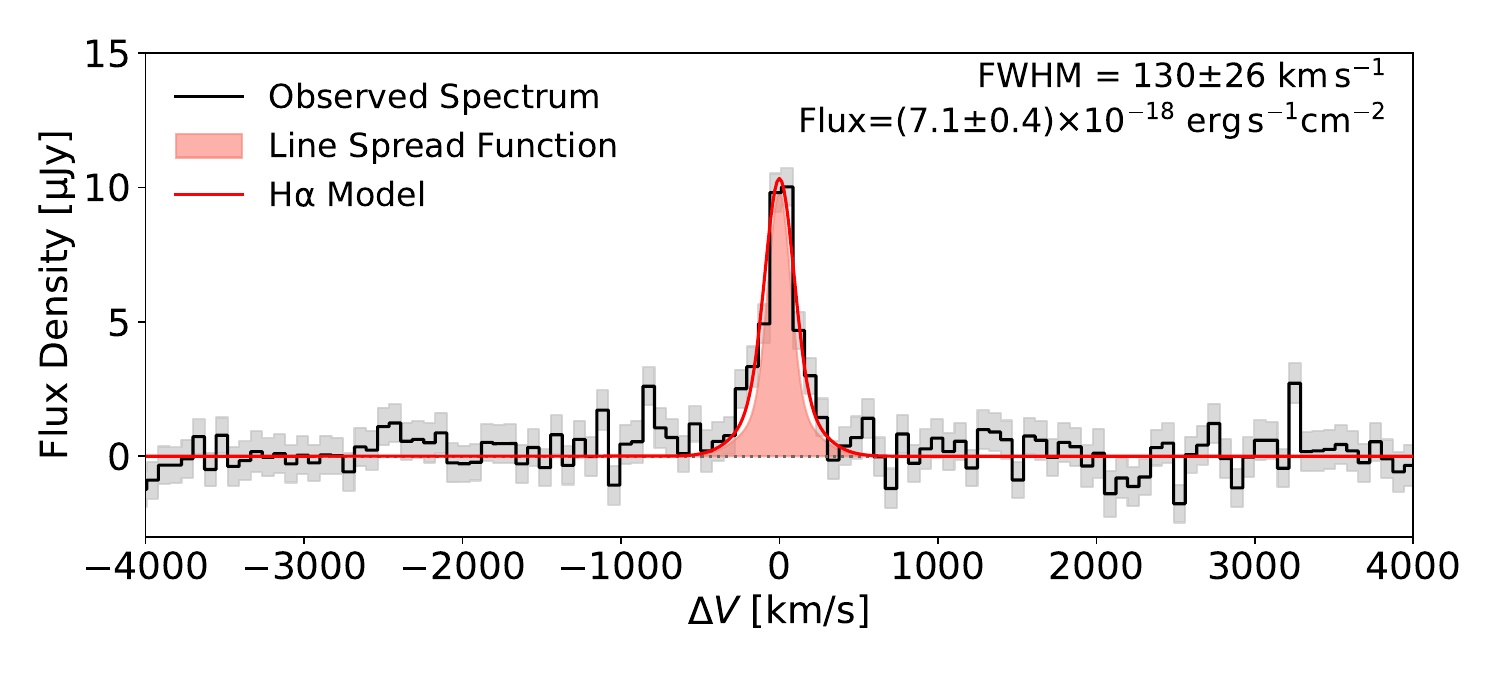}
\includegraphics[width=0.49\linewidth]{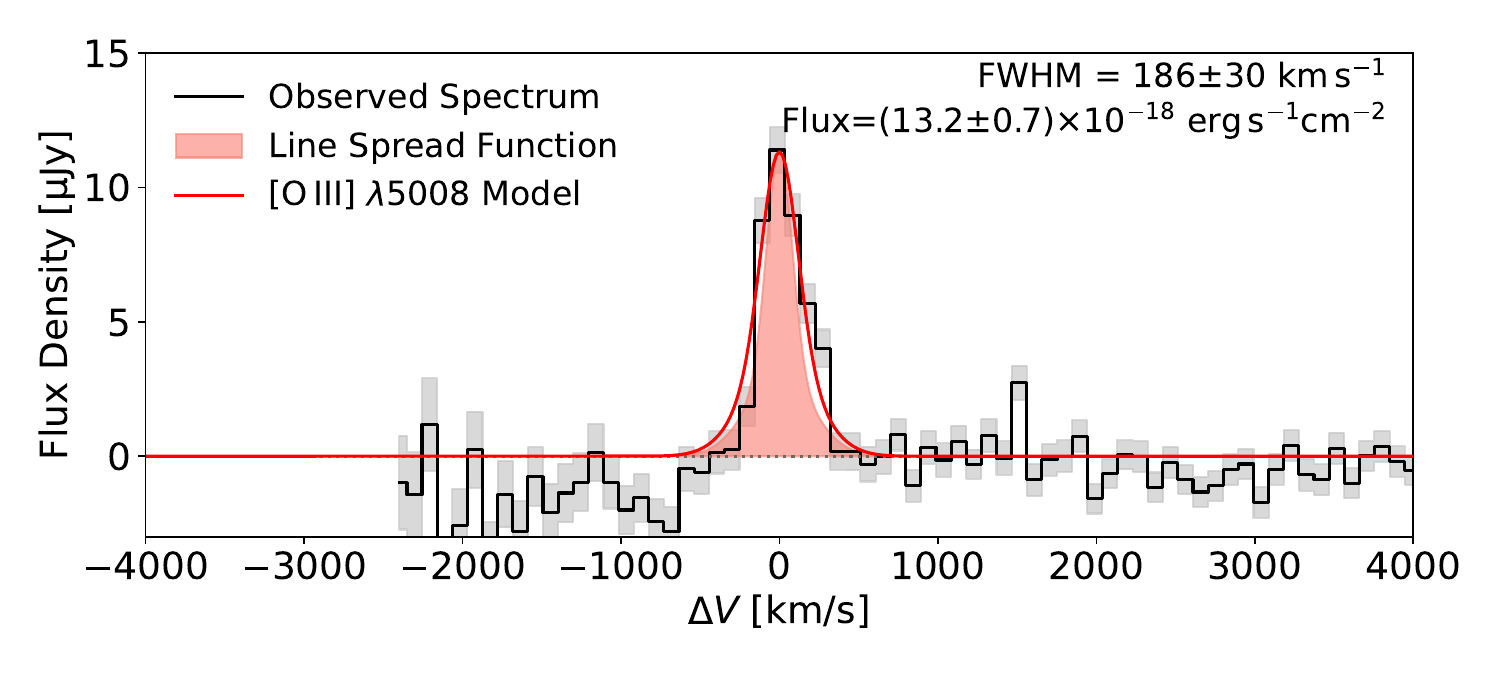} 
\caption{
NIRCam grism spectra of the H$\alpha$ (left; with FRESCO F444W) and [\ion{O}{3}] $\lambda$5008 (right; with CONGRESS F356W) emission lines of AT 2023adya's host, shown as solid black lines with uncertainties in gray shades.
Line spread functions (Sun, F.\ et al., in prep.) at line centroids are shown as shallow red shades, and best-fit Gaussian profiles convolved with the line spread functions are shown as the solid red lines. Both lines do not show hints of broad components.
Note that the [\ion{O}{3}] $\lambda$5008 line is at the blue edge of the F356W transmission and the continuum / background subtraction is not optimal.
}
\label{fig:host_spectra}
\end{figure*}

AT 2023adya's host has not been previously identified as an AGN candidate \citep[]{Lyu2022, Lyu2024}, but this is not unexpected due to its relatively high redshift and limited wavelength coverage. A broad H$\alpha$ emission line generated within the AGN BLR is a Type 1 AGN signature. As shown in the left side of Figure \ref{fig:host_spectra}, we fit the host's H$\alpha$ emission with a single Gaussian component considering the grism line spread function calibrated by Sun, F.\ et al.\ (in preparation), and it does not exhibit broad H$\alpha$ emission. Rather, the FWHM of the H$\alpha$ line is only 130$\pm$26 km/s (the H$\alpha$ line flux is 7.1$\pm$0.4$\times$10$^{-18}$ erg\,s$^{-1}$\,cm$^{-2}$). This F444W NIRCam/Grism spectrum showing narrow H$\alpha$ was taken approximately one observer-frame week after the Epoch1 NIRCam imaging. The right side of Figure \ref{fig:host_spectra} shows a single-component fit to the host's [\ion{O}{3}] $\lambda$5008 emission line, which has a FWHM of 186$\pm$30 km s$^{-1}$ and a line flux of 1.32$\pm$0.07$\times$10$^{-17}$ erg\,s$^{-1}$\,cm$^{-2}$. This F356W NIRCam/Grism spectrum was taken at the time of Epoch2.

It is difficult to exclude the existence of a faint broad-line component based on the limited-depth NIRCam/Grism spectrum. In fact, \citet{zhang2025} mentions two sources that exhibit no broad features in NIRCam/Grism spectra but are reported to show broad line features in NIRSpec medium-resolution grating data \citep{maiolino2024}. \citet{zhang2025} suggests that broad line searches in NIRCam/Grism data are effective down to $\sim$26~mag, and the Epoch1 F444W magnitude of AT 2023adya's host is slightly fainter than that limit ($\sim$26.4~mag). 

To test whether or not we would expect to detect broad H$\alpha$ in the NIRCam/Grism data for various Type 1 AGN configurations, we roughly estimated the expected H$\alpha$ line flux based on the observed F356W flux density using Equation 1 from \citet{greene2005}. At $z$\,$=$\,5.274, F356W corresponds to rest-frame 5674~\AA, which we approximated as 5100~\AA. We tested two different scenarios: (1) AT 2023adya is a Type 1 AGN whose continuum is AGN-dominated, meaning that the observed Epoch1 F356W flux density ($\sim$\,138~nJy) is entirely from the AGN, and (2) AT 2023adya is a Type 1 AGN whose continuum is dominated by star formation, meaning that the AGN only contributes to the observed change in F356W brightness ($\sim$\,23~nJy). 

In the first scenario (pure Type 1 AGN), the AGN flux density of $\sim$138 nJy corresponds to an expected H$\alpha$ line flux of $\sim$4\,$\times$\,10$^{-18}$ erg s$^{-1}$ cm$^{-2}$, which would constitute $\sim$60\% of the observed H$\alpha$ line flux. Given this significant contribution to the total line flux, we would expect to see a broad component in the H$\alpha$ line if AT 2023adya were a pure Type 1 AGN. Therefore, our lack of broad H$\alpha$ detection means that it is unlikely that AT 2023adya is a Type 1 AGN whose continuum is dominated by the AGN. Conversely, if AT 2023adya is a Type 1 AGN whose continuum is dominated by star formation, the $\sim$23~nJy from the AGN would constitute an H$\alpha$ line flux of only $\sim$6.5\,$\times$\,10$^{-19}$ erg s$^{-1}$ cm$^{-2}$, which is roughly ten times less than the measured H$\alpha$ line flux. In this case, it is unlikely that we would detect the broad component of H$\alpha$, especially if the line is significantly broadened, given NIRCam/Grism's limited sensitivity. Therefore, we cannot exclude the possibility that AT 2023adya is a star formation-dominated Type 1 AGN.


\subsubsection{Type 2 AGN} \label{subsubsec:type2_agn}

According to the simple AGN Unified Model \citep{antonucci1993}, the central engines and broad line regions of Type 2 AGN are obscured by toroidal dust distributions. These dusty tori highly suppress optical variations, so we expect Type 2 AGN to exhibit little to no optical variability. Even if we could see the AGN continuum through scattered light, the non-variable stellar emission from the host would dominate and dilute the variability of the scattered AGN light. \citet{Barth2014} validated these predictions by searching for optical \textit{g}-band variability among the luminous Type 2 AGN in the Sloan Digital Sky Survey (SDSS) Stripe82 field. They found that $\sim$90\% of the Type 2 AGN did not show optical variability, and of the 10\% that did show variability, most if not all of them are misclassified Type 1 AGN. Notably, however, the misclassified Type 1 AGN in \citet{Barth2014} exhibited stronger variability than AT 2023adya. 

There is a unique subsample of Type 2 AGN that exhibit similar levels of optical variability as AT 2023adya. These are the so-called ``true Type 2” or ``naked Type 2” AGN, which are unobscured AGN (i.e., viewed face-on) that intrinsically lack a BLR \citep{hawkins2004}. Previous AGN variability surveys such as those conducted by \citet{cartier2015} and \citet{sanchez2017} have reported true Type 2 AGN candidates. Furthermore, \citet{Barth2014} were unable to exclude the true Type 2 AGN scenario for a small subset of their variable Type 2 AGN sample, although they disfavor this explanation. We are also unable to exclude the possibility that AT 2023adya is a true Type 2 AGN, but we do not favor this possibility due to the rarity of these objects \citep{Barth2014}.

\subsection{Tidal Disruption Event} \label{subsec:tde}

A TDE occurs when a supermassive black hole (SMBH) shreds apart a star that passes within its tidal radius, generating a luminous accretion flare \citep[]{Rees1988, Evans1989, Ulmer1999}. TDEs are expected to be abundant at high redshifts due to the high central stellar densities of high redshift galaxies and the slightly lower SMBH masses \citep{Karmen2025}. \citet{Inayoshi2024} find that deep \textit{JWST} surveys can discover TDEs occurring in moderately obscured AGN (e.g., with typical attenuation levels of $A_{\rm V} \sim 3$ mag, as seen in LRDs) up to $z$\,$\approx$\,4--7 even with small survey areas. These observations enable the exploration of the mass distribution of SMBHs at the low-mass end, where their abundance is sufficiently high for detection. \citet{Pfister2021} uses hydrodynamical simulations to show that the TDE rate can be enhanced at early cosmological times, and that TDEs could be a viable growth mechanism for massive black hole seeds. Additionally, \citet{Karmen2025} reports the discovery of a high-redshift TDE candidate in the COSMOS-Web field, identified using novel color and morphology selection criteria. This further suggests enhanced TDE rates at high-redshift relative to the local rate. 

Despite these predictions, however, it is difficult to explain AT 2023adya as a TDE. TDEs are extremely UV-bright, and yet AT 2023adya exhibits no change in rest-frame UV brightness. 
Therefore, if AT 2023adya is a TDE, it must be embedded in the dusty nucleus of a galaxy \citep[e.g.][]{Masterson2024}.  
Although it is possible to produce a dust-obscured TDE SED that is red enough to 
explain the F356W detection and the F090W/F115W non-detections of AT 2023adya, producing the observed F356W magnitude of $\sim$28\,mag for a $z$\,$=$5 TDE is not straightforward.  
For example, the $z$\,$=$\,5 dust-obscured TDE SED model of \citet{Inayoshi2024} with M$_{\rm SMBH}=10^5$\,\Msun\ and $A_{\rm V}\gtrsim 3$\, mag in their Figure~8 can reproduce the required SED color and match the high detection rate inferred by the small survey area of $\sim38~{\rm arcmin}^2$ (corresponding to $\sim100~{\rm deg}^{-2}{\rm yr}^{-1}$, see their Figure~5). However, this model remains $\sim$1 mag fainter ($\sim$29~mag) than AT 2023adya's F356W detection even at peak brightness. 
While a more luminous TDE model with a luminosity exceeding $\sim10^{45}~{\rm erg~s}^{-1}$ (observationally rarer in the nearby universe; see Table~1 of \citealt{Gezari2021}) or a more massive SMBH could match the observed brightness, both options would imply a lower event rate and space density, making detection within our survey less likely \citep[e.g.,][]{Inayoshi2024}. 

\section{Conclusion} \label{sec:conclusion}

We discovered AT 2023adya by differencing F356W NIRCam imaging in GOODS-N taken one observer-frame year apart by the JADES and CONGRESS programs. AT 2023adya is thus far the highest-redshift transient/variable source showing a change in photometric brightness discovered with \textit{JWST}, residing in a $z_{\mathrm{spec}}$\,$=$\,5.274 galaxy which faded from $m_{\mathrm{F356W}}$\,$=$\,26.05$\pm$0.02~mag to 26.24$\pm$0.02~mag (+0.19~mag) over approximately two rest-frame months. The F356W difference image shows a clear residual signal ($m_{\mathrm{F356W}}$\,$=$\,28.01$\pm$0.17~mag). AT 2023adya does not, however, exhibit a rest-frame UV brightness change in the F090W or F115W filters. While we cannot determine the true nature of AT 2023adya, we explore various sources that may create the observed brightness change, including an SN explosion, AGN, and a TDE.

\begin{itemize}

    \item Any CCSN subtype can explain AT 2023adya's F356W magnitude, although each CCSN subtype except for SNe IIn would need to be brighter than average. Although we cannot exclude the possibility that AT 2023adya is an SN Ia based on its observed brightness, the bright CCSN scenario is more likely because CCSN rates are expected to dominate over SN Ia rates at high redshift.

    \item We cannot rule out the possibility that AT 2023adya is a Type 1 AGN whose continuum is dominated by star formation. In this case, it is likely that the broad H$\alpha$ component would be too faint for detection in the F444W NIRCam/Grism spectrum. Additionally, we cannot rule out the possibility that AT 2023adya is a ``true" or ``naked" Type 2 AGN, but we do not favor this possibility due to the rarity of these objects.
    

    \item It is unlikely that AT 2023adya is a TDE due to the low event rate of TDEs matching the observed photometry (i.e., with sufficient dust to obscure the rest-frame UV emission and sufficient optical luminosity to match the F356W photometry).
    
\end{itemize}

Although we cannot determine AT 2023adya's true nature with the available data, this discovery is significant because it represents the first piece of photometric evidence that \textit{JWST} can robustly detect transient/variable sources at $z$\,$>$\,5. It is notable that the JTS discovered no such sources at $z$\,$>$\,5 in GOODS-S. Had a source similar to AT 2023adya been in the JADES Deep Field within GOODS-S, the JTS selection criteria likely would have identified it. The lack of $z$\,$>$\,5 transient and variable sources in the JTS sample indicates the rarity of discovering such sources at high redshift when the temporal baseline between observations is only one observer-frame year. To continue building the $z$\,$>$\,5 transient and variable source sample, which may include exotic sources such as Population III SNe, we will conduct multi-epoch multi-filter \textit{JWST} observations with baselines longer than one observer-frame year through Cycle-4 \textit{JWST} Program 8060 (PI: Egami).


\begin{center}
    \textbf{Acknowledgements}
\end{center}
We thank the anonymous referee whose comments and
suggestions led to significant improvements in the paper. This work is based on observations made with the NASA/ESA/CSA James Webb Space Telescope. The data were obtained from the Mikulski Archive for Space Telescopes at the Space Telescope Science Institute, which is operated by the Association of Universities for Research in Astronomy, Inc., under NASA contract NAS 5-03127 for JWST. These observations are associated with program \#1181, 1895 and 3577. The HLF, FRESCO, JADES, and CONGRESS data used in this paper can be found on MAST \citep{hlf2015, fresco2023, jades2024, congress2025}. Support for program \#3577 was provided by NASA through a grant from the Space Telescope Science Institute, which is operated by the Association of Universities for Research in Astronomy, Inc., under NASA contract NAS 5-03127. The authors acknowledge the FRESCO team for developing their observing program with a zero-exclusive-access period.

A.J.B. acknowledges funding from the “FirstGalaxies” Advanced Grant from the European Research Council (ERC) under the European Union’s Horizon 2020 research and innovation program (Grant agreement No. 789056). K.I. acknowledges support from the National Natural Science Foundation of China (12073003, 12003003, 11721303, 11991052, and 11950410493) and the China Manned Space
Project (CMS-CSST-2021-A04 and CMS-CSST-2021-A06). S.M. acknowledges support from the Research Council of Finland project 350458. D.P. acknowledges support by the Huo Family Foundation through a P.C. Ho PhD Studentship. E.E., B.D.J., G.R., and B.E.R. acknowledge support from the JWST/NIRCam contract to the University of Arizona NAS5-02015. B.E.R. also acknowledges support from the JWST Program 3215. S.T. acknowledges support by the Royal Society Research Grant G125142. The research of C.C.W. is supported by NOIRLab, which is managed by the Association of Universities for Research in Astronomy (AURA) under a cooperative agreement with the National Science Foundation. The authors acknowledge use of the lux supercomputer at UC Santa Cruz, funded by NSF MRI grant AST 1828315.

\facilities{\textit{JWST}, \textit{HST}}

\software{astropy \citep{astropy2013, astropy2018, astropy2022}, photutils \citep{Bradley2024Astropy/photutils:1.12.0}}

\bibliography{references}{}
\bibliographystyle{aasjournal}

\end{document}